\documentclass{appolb}
\usepackage{graphicx}

\begin{document}

\title{Photon induced processes 
in semi-central \\ nucleus-nucleus collisions
\thanks{presented at the Epiphany 2019 workshop}
}

\author{Antoni Szczurek
\address{Institute of Nuclear
Physics, Polish Academy of Sciences, Radzikowskiego 152, PL-31-342
Krak{\'o}w, Poland,\\
University of Rzesz\'ow, PL-35-959 Rzesz\'ow, Poland
}
}

\maketitle

\begin{abstract}
We calculate total and differential cross sections
for $J/\psi$ photoproduction in ultrarelativistic lead-lead collisions
at the LHC energy $\sqrt{s_{NN}}=2.76$ TeV.
We use a simple model based on vector dominance
picture and multiple scattering of the hadronic ($c \bar c$) state 
in a cold nucleus.
In our analysis we use Glauber formulae for calculating 
$\sigma_{tot,J/\psi Pb}$ which is a building block of our model.
We compare our UPC results with ALICE data.
For semi-central collisions a modification of the photon flux is necessary. 
We discuss how to effectively correct photon fluxes for geometry
effects.
We try to estimate the cross sections for different centrality bins
and for $J/\psi$ mesons emitted in forward rapidity range ($2.5<y<4$)
corresponding to the ALICE experimental results. 

We discuss similar analysis for dilepton production in 
ultrarelativistic heavy-ion collisions at very low pair transverse 
momenta,  $P_T\leq 0.15$\,GeV.  
We investigate the interplay of thermal radiation with
photon annihilation processes, $\gamma \gamma \to l^+ l^-$, 
due to the coherent electromagnetic fields of the colliding
nuclei. For the thermal radiation, we employ the emission from the QGP 
and hadronic phases with in-medium vector spectral functions.          
We first verify that the combination of photon fusion, thermal 
radiation and final-state hadron decays gives a fair description of 
the low-$P_T$ invariant-mass as well as $P_T$ distributions as 
measured recently by the STAR collaboration in 
$\sqrt{s_{NN}}$=200\,GeV Au+Au collisions for different centralities.
The coherent contribution dominates in peripheral collisions, 
while thermal radiation shows a significantly stronger increase with
centrality. 
We also provide predictions for the ALICE experiment at the LHC. 
The resulting excitation function reveals a nontrivial interplay 
of photoproduction and thermal radiation.  
\end{abstract}


\section{Introduction}

The $J/\psi$ production in heavy-ion collisions was considered 
as a flag example of quark-gluon plasma.
Simultaneusly $J/\psi$ was studied in ultraperipheral collisions
when nuclei do not touch. In this case
a coherent photon, which couples to one of the colliding nuclei,
fluctuates into a virtual $J/\psi$ or $c \bar c$ pair which then is
produced as $J/\psi$ meson in the final state. Till recently
it was not discussed what happens to the photoproduction processes
when nuclei collide and presumably quark-gluon plasma is created.
Quite recently the ALICE collaboration observed 
$J/\psi$ with very small transverse momenta in peripheral 
and semi-central collisions \cite{ALICE2016}.
This was interpreted in \cite{KS2016} as effect of photoproduction
mechanism which is active also in such a case.

Recently the STAR collaboration observed also enhanced production of 
dielectron pairs with small transverse momenta \cite{Adam:2018tdm}.
We showed very recently \cite{KRSS2019} that this may be interpreted as 
$\gamma \gamma \to e^+ e^-$ processes (with coherent photons) even 
in the semi-central collisions.

In this presentation we discuss what happens with the coherent photon 
induced processes in the semi-central collisions. Two examples 
are presented:\\
(a) photoproduction of $J/\psi$ meson,\\
(b) production of dilepton pairs.

\section{Sketch of the formalism}

\subsection{$J/\psi$ production}

\begin{figure}[h!]
\centering
\includegraphics[width=4cm]{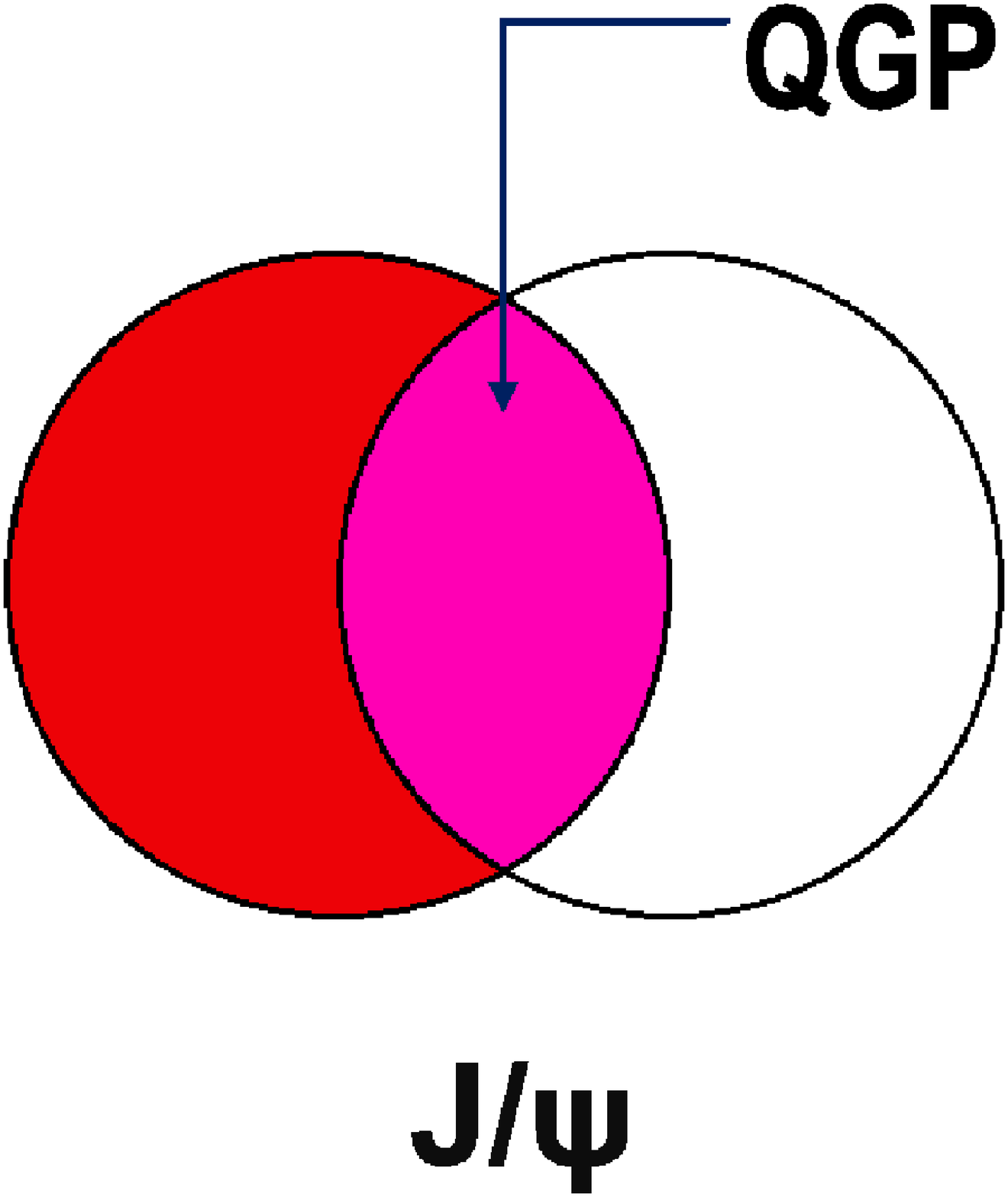}\qquad
\includegraphics[width=4cm]{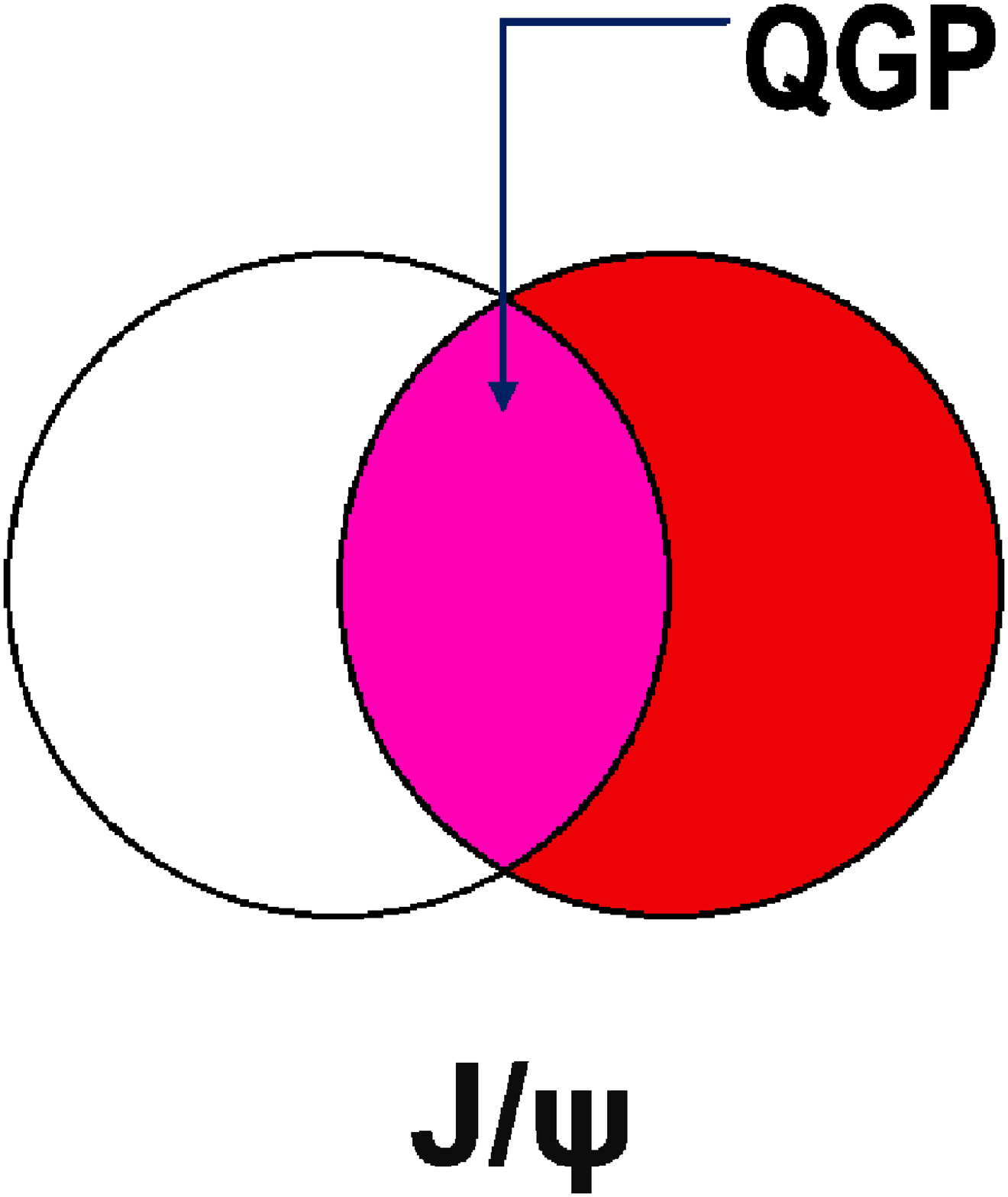}
\caption{Emission of $J/\psi$ - a picture in the plane $x,y$
  perpendicular to the collision axis
    ($z$). In the gray area quark-gluon plasma is created.}
\label{fig:impact_parameter_jpsi}
\end{figure}

We start from presentation of the situation for $J/\psi$ meson
production in semi-central collisions. In
Fig.\ref{fig:impact_parameter_jpsi} we show the situation in the
impact parameters space. Either first or second ion emits a photon.
The corresponding hadronic fluctuation rescatters then in the second
or first nucleus, respectively.

\begin{figure}[!h]
\includegraphics[scale=0.55]{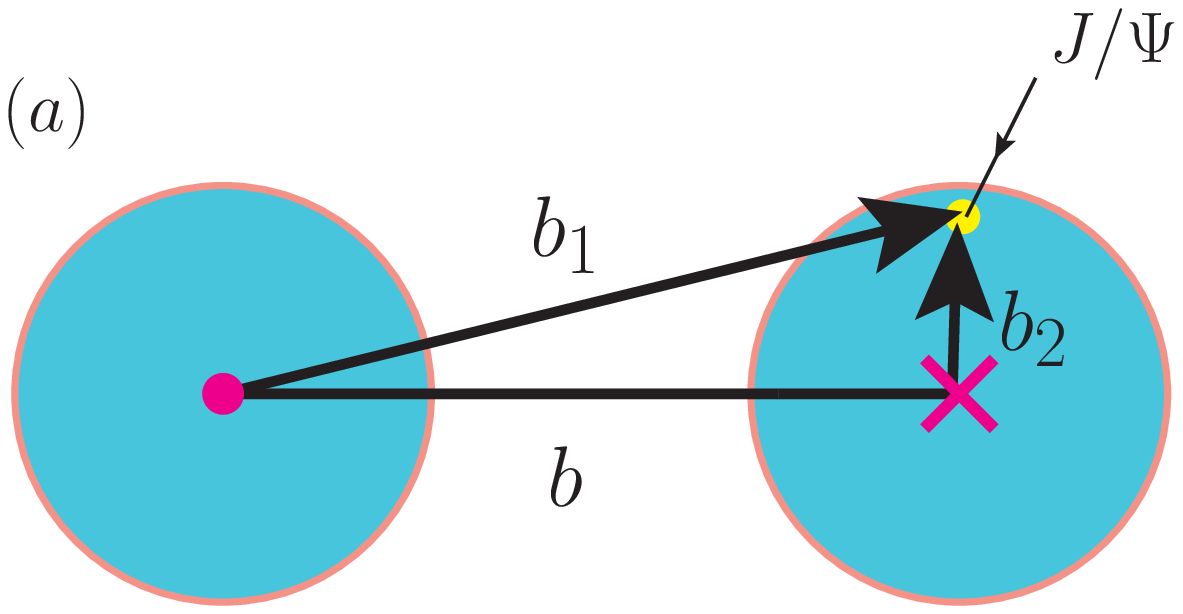}
\includegraphics[scale=0.55]{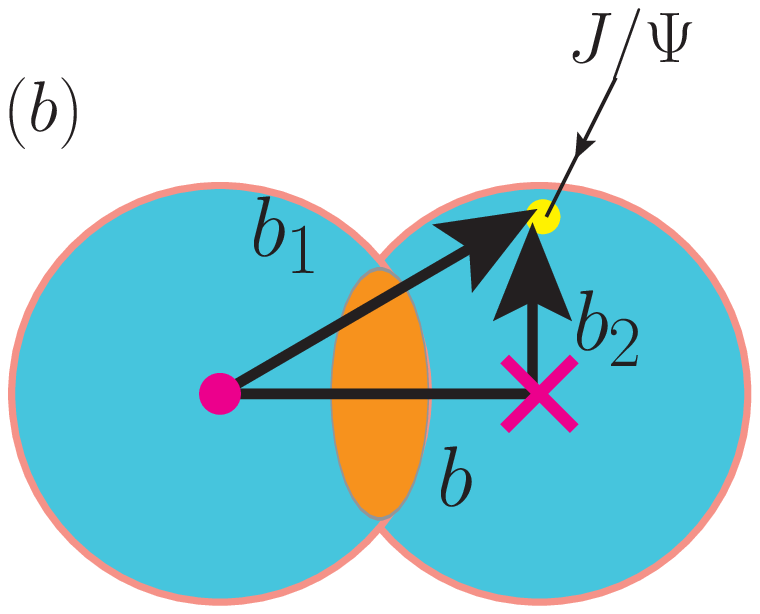}
\caption{Impact parameter picture of the production of the $J/\psi$
meson for ultraperipheral (left panel)
and for semi-central (right panel) collisions.
It is assumed here that the first nucleus is the emitter of the photon
which rescatters then in the second nucleus being a rescattering medium.
}
\label{fig:general_situation}
\end{figure}

In Fig.\ref{fig:general_situation} we compare situation for
ultra-peripheral (left) and semi-central (right) collisions.
Is $J/\psi$ created before nuclear collision ?
If yes, it would be easy to melt $J/\psi$ in the quark-gluon plasma
(orange).

\begin{figure}[!h]
\centering
    \includegraphics[scale=0.20]{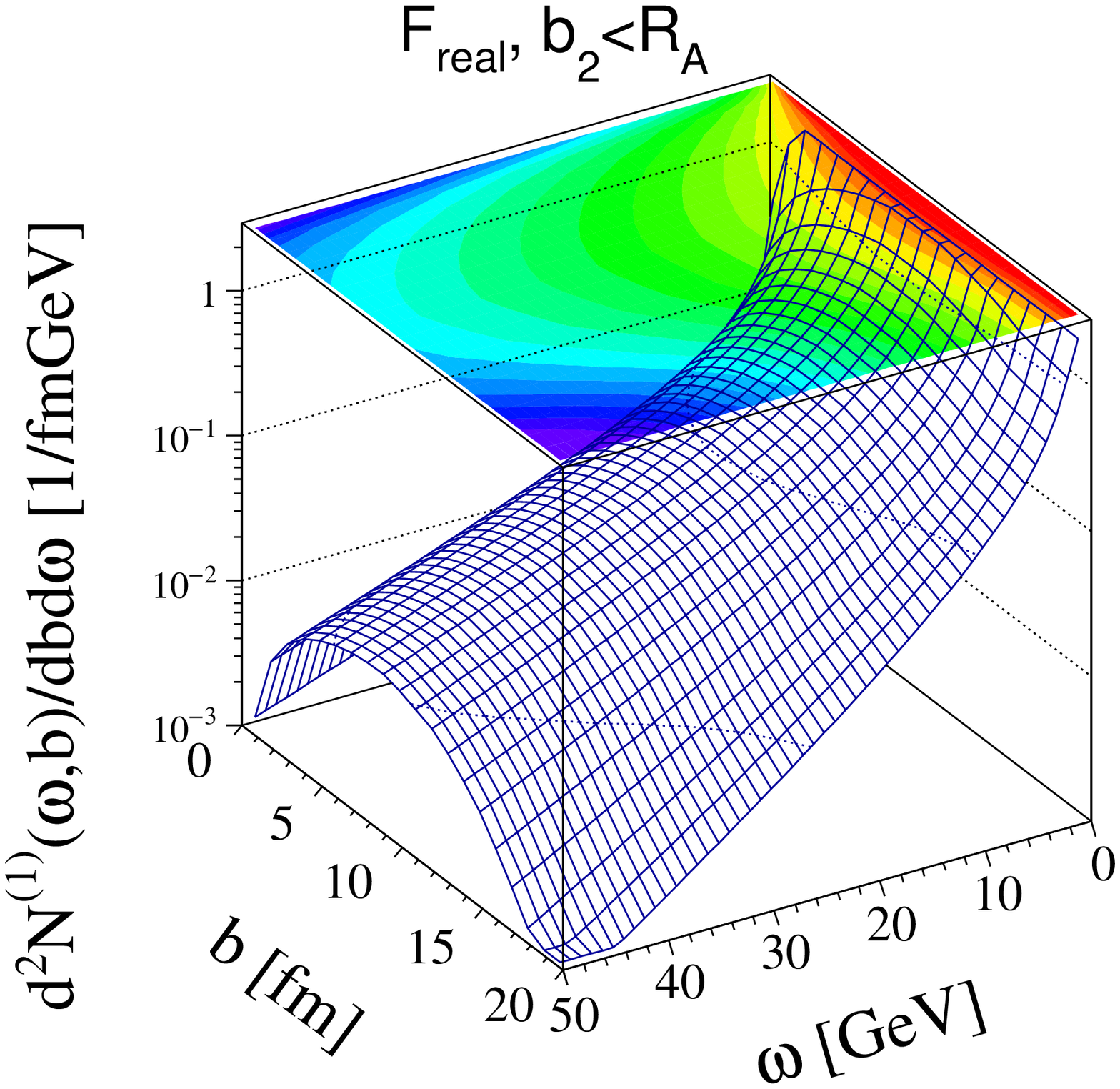}
    \includegraphics[scale=0.20]{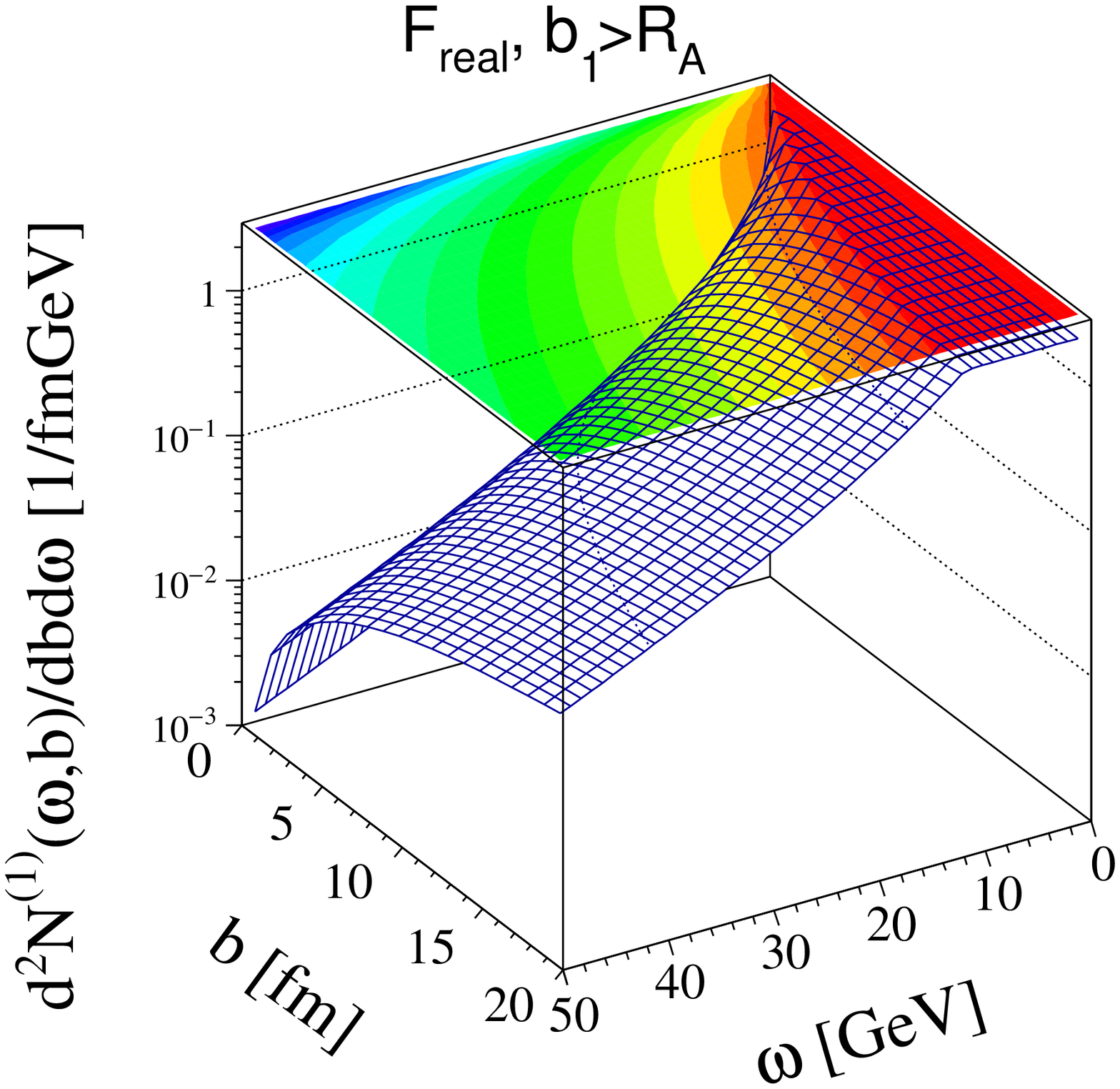}
    \includegraphics[scale=0.20]{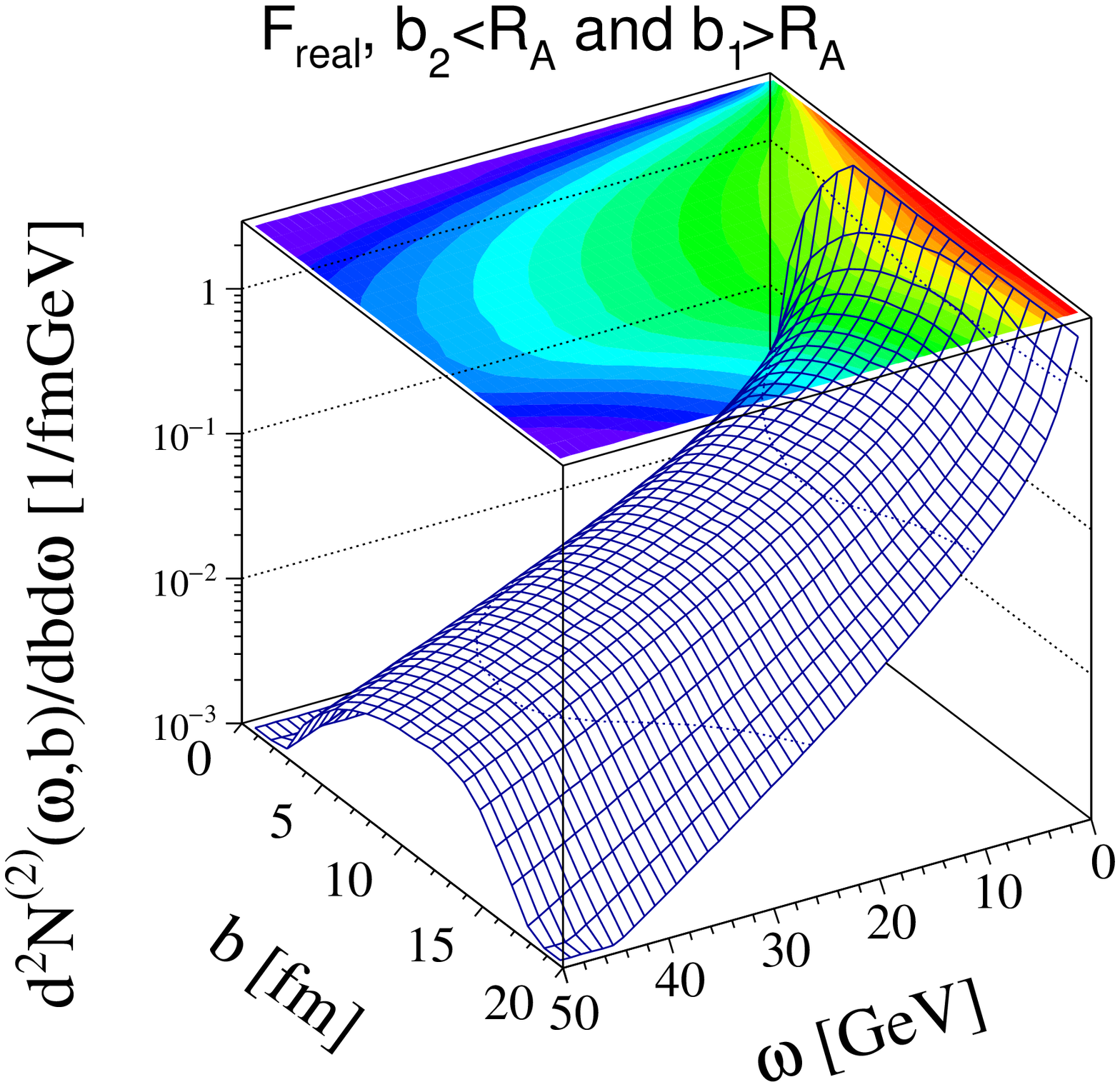}
\caption{Two-dimensional distributions of the photon flux 
in the impact parameter $b$ and in the energy of photon $\omega$
for three different conditions (more in the text).}
\label{fig:d2N_dbdomega}
\end{figure}

The details how to calculate cross section were exposed in
\cite{KS2016}. We proposed that compared to UPC collisions
one should modify photon flux factors.
The effective photon flux which includes the geometrical aspects 
can be formally expressed through 
the real photon flux of one of the nuclei and effective 
strength for the interaction of the photon with the second nucleus
\begin{equation}
 N^{(1)}\left(\omega_1,b\right) = \int N\left(\omega_1,b_1\right)
\frac{\theta(R_A - (|\overrightarrow{b_1}-\overrightarrow{b}|))}{\pi R_A^2} \mathrm{d}^2 b_1 
\; ,
\label{eq:d2N/domegadb}
\end{equation}
where $\overrightarrow{b_1} = \overrightarrow{b} + \overrightarrow{b_2}$. 
The extra $\theta(R_A - (|\overrightarrow{b_1}-\overrightarrow{b}|))$ factor
ensures collision when the photon hits
the nucleus-medium. For the photon flux in the second nucleus
one needs to replace 1$\to$2 (and 2$\to$1).
For large $b \gg R_A+R_B$: $N^{(1)}\left(\omega_1,b\right) \approx N(\omega_1,b)$.
For small impact parameters this approximation is,
however, not sufficient. 
This has some consequences also for ultraperipheral
collisions, which will be discussed somewhat later in this section.
Since it is not completely clear what happens in the region 
of overlapping nuclear densities we suggest another approximation 
which may be considered rather as lower limit. 
In this approximation we integrate the photon flux of 
the first (emitter) nucleus only over this part of the second 
(medium) nucleus which does not collide with the nucleus-emitter
(some extra absorption may be expected in the tube
of overlapping nuclei). 
This may decrease the cross section for more central collisions. 
In particular, for the impact parameter $b=0$ the resulting vector meson 
production cross section will fully disappear by the construction.
In the above approximation the photon flux can be written as: 
\begin{equation}
N^{(2)}\left(\omega_1,b\right) = \int N\left(\omega_1,b_1\right)
\frac{\theta(R_A - (|\overrightarrow{b_1}-\overrightarrow{b}|)) \times \theta(b_1 - R_A)}
{\pi R_A^2} \mathrm{d}^2 b_1  \; .
\label{eq:d2N/domegadb_second}
\end{equation}
In our calculation we use the following generic formula for
calculating the photon flux for any nuclear form factor $F$
%
\begin{equation}
N(\omega,b) = \frac{Z^2 \alpha_{em}}{\pi^2} \left| \int u^2 J_1\left(u\right)
\frac{F\left( \frac{\left( \frac{\omega b}{\gamma} \right)^2+u^2}{b^2} \right)}{\left(\frac{\omega b}{\gamma} \right)^2+u^2} \right|^2 \;.
\label{eq:N_omegab}
\end{equation}
The fluxes including different limitations are shown in 
Fig.\ref{fig:d2N_dbdomega}.

\subsection{Dilepton production}

\begin{figure}[h!]
\centering
\includegraphics[width=5cm]{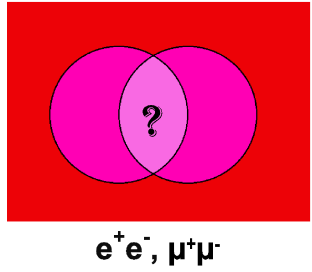}
\caption{Dilepton production - a picture in the plane $x,y$
  perpendicular to the collision axis ($z$).
The area with the question mark is the region where quark-gluon plasma
is created.
}
\label{fig:impact_parameter_epem}
\end{figure}

The general picture in the impact parameter space is shown in
Fig.\ref{fig:impact_parameter_epem}.
 
There are general questions one can add concerning the considered reaction.
Is $e^+ e^-$ created before nuclear collision ?
Can plasma/spectators distort distributions of leptons ?
Here a big fraction of events can be produced outside (in the impact
parameter space) of both colliding nuclei (the dark area). Therefore 
not big difference of the cross sections between UPC and non-UPC is expected.

The main ingredient for the photon-photon fusion mechanism is 
the flux of photons for an ion of charge $Z$ moving
along impact parameter with the relativistic parameter $\gamma$. 
With the nuclear charge form factor $F_{\rm em}$ as an input the flux
can be calculated as~\cite{Bertulani:1987tz, Baur:2001jj}
\begin{eqnarray}
N(\omega,b) 
&&= {Z^2 \alpha_{\rm EM} \over \pi^2} 
\Big| \int_0^\infty  dq_t {q_t^2   F_{\rm em}(q_t^2 + {\omega^2 \over \gamma^2} )   
	\over q_t^2 + {\omega^2 \over \gamma^2} } J_1(b q_t) \Big|^2\, ,  
\label{eq:WW-flux}
\end{eqnarray}
where $J_1$ is a Bessel function, $q_t$ is photon transverse momentum
and $\omega$ is photon energy. 
We calculate the form factor from the Fourier transform of the 
nuclear charge density.

The differential cross section for dilepton ($l^+ l^-$) production via 
$\gamma \gamma$ fusion at fixed impact parameter of a nucleus
nucleus collision can then be written as
\begin{eqnarray}
{d \sigma_{ll} \over d\xi d^2 b } =  
\int d^2b_1 d^2b_2 \, \delta^{(2)}(\vec{b} - \vec{b}_1 - \vec{b}_2) N(\omega_1,b_1) N(\omega_2,b_2) 
{d \sigma(\gamma \gamma \to l^+ l^-; \hat s) \over d p_t^2} \ ,
\end{eqnarray}
where the phase space element is $d\xi = dy_+ dy_- dp_t^2$ with $y_\pm$, $p_t$ and $m_l$ the single-lepton 
rapidities, transverse momentum and mass, respectively,
and
\begin{eqnarray}
\omega_1 = {\sqrt{p_t^2 + m_l^2} \over 2} \, ( e^{y_+} + e^{y_-} ) \  , \  
\omega_2 = {\sqrt{p_t^2 + m_l^2} \over 2} \, ( e^{-y_+} + e^{-y_-} ) \ , \ \hat{s} = 4 \omega_1 \omega_2 \ .
\end{eqnarray}
As can be seen from Eq.(\ref{eq:WW-flux}), the transverse momenta,
$q_t$, of the photons have been 
integrated out, and in this approximation dileptons are produced 
back-to-back in the transverse plane. 

In UPCs the incoming nuclei do not touch, i.e. no strong interactions
occur between them. In this case one usually imposes the constraint 
$b > 2 R_A$ when integrating over impact parameter.
In semi-central collisions we lift this restriction allowing the nuclei 
to collide.

An exact calculation of the pair-$P_T$ dependence is, in general,  
rather involved.
In Ref.\cite{KRSS2019} we performed a simplified calculation
using $b$-integrated transverse momentum dependent photon fluxes, 
\begin{equation}
{dN(\omega,q^2_t) \over d^2 \vec{q}_t} = {Z^2 \alpha_{EM} \over \pi^2} 
{q_t^2 \over [q_t^2 + {\omega^2 \over \gamma^2}]^2} F^2_{\rm em}(q_t^2 + {\omega^2 \over \gamma^2} ).   
\end{equation}
%
The $P_T$ distribution is then obtained 
as the convolution of two transverse momentum dependent photon fluxes 
with the elementary $\gamma \gamma \to e^+ e^-$ cross section,
\begin{equation}
\frac{d \sigma_{ll}}{d^2 \vec{P}_{T}} 
= \int {d\omega_1 \over \omega_1} {d \omega_2 \over \omega_2} 
d^2 \vec{q}_{1t} d^2\vec{q}_{2t} \frac{dN(\omega_1,q_{1t}^2)}{d^2 \vec{q}_{1t}}
  \frac{dN(\omega_2,q_{2t}^2)}{d^2 \vec{q}_{2t}}
\delta^{(2)}(\vec{q}_{1t}+\vec{q}_{2t}-\vec{P}_{T}) 
{\hat \sigma}(\gamma \gamma \to l^+ l^-)\Big|_{\rm cuts} \; ,
\label{pairtransversemomentum_distribution}
\end{equation}
The resulting  shape of integrated cross section is then renormalized
to the previously obtained cross section obtained in the collinear
approximation for a given centrality class.

\section{Selected results}

\begin{figure}[!h]
\centering
\includegraphics[scale=0.35]{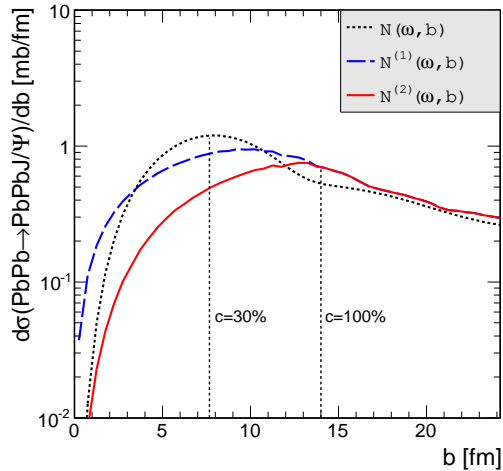}
\caption{Differential cross section for photoproduction
of $J/\psi$ meson as a function of impact parameter 
for $\sqrt{s_{NN}}=2.76$~TeV. 
Different lines correspond to different approximations:
dotted - standard UPC approach, 
dashed - first approximation/correction (upper limit),
solid - second approximation/correction (lower limit). 
Here realistic (charge) nuclear form factor was used.
For reference we show vertical lines corresponding
to centralities $c=30\%$ and $c=100\%$.
}
\label{fig:dsig_db}
\end{figure}

In Fig.~\ref{fig:dsig_db} we show the nuclear cross section for $J/\psi$
production as a function of the impact parameter also for
$b < R_A + R_B$ i.e. for the semi-central collisions.
We show results for a broad range of impact parameter
($0<b<R_A+R_B$). However, the application of our approach
for very small $b$ is not obvious.

The different lines correspond to different approximations of photon fluxes
within our approach as described in the figure caption.
The dashed and solid lines represent
upper and lower limit for the cross section.
At larger values of impact parameter $b$ the cross sections obtained
with the different fluxes practically coincide.
At $b < R_A + R_B$ the different approximations give quite different results.
The standard approach in the literature for UPC
when naively applied to the semi-central collisions overestimates 
the cross section.

\begin{figure}[!h]
\centering
\includegraphics[scale=0.35]{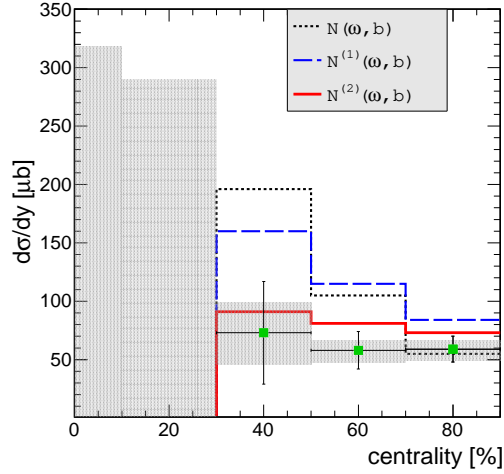}
\caption{$\Delta \sigma / \Delta y$ cross sections for different 
centrality bins. Theoretical results for different
models of the photon flux are compared with the
ALICE data \cite{ALICE2016}.
The shaded area represents the experimental uncertainties.}
\label{fig:flux_hist}
\end{figure}

We summarize our calculations in \cite{KS2016}
in Fig.~\ref{fig:flux_hist}.
We present both statistical and systematic error bars (shaded area).
We show our results starting from centralities bigger than $30 \%$.
As discussed above we do not trust our results for lower centralities.
In addition, the ALICE Collaboration could not extract actual values
of the cross section for the two lowest centrality bins. 
The results for standard photon flux exceed the ALICE data.
Rather good agreement with the data is achieved 
for the $N^{(2)}$ photon flux obtained with the realistic
nucleus form factor. 

\begin{figure}[!t]
	\includegraphics[scale=0.35]{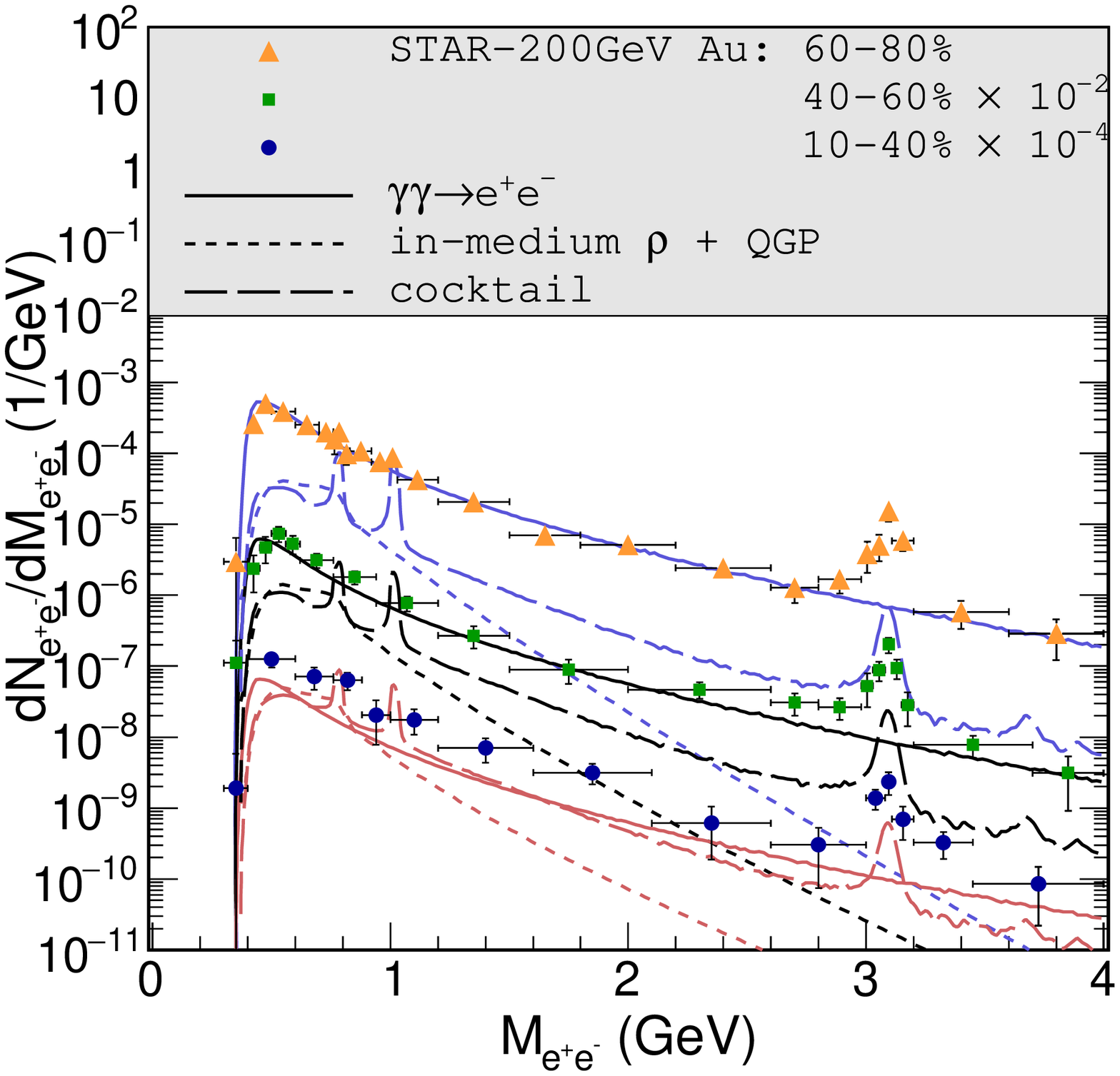}
	\includegraphics[scale=0.35]{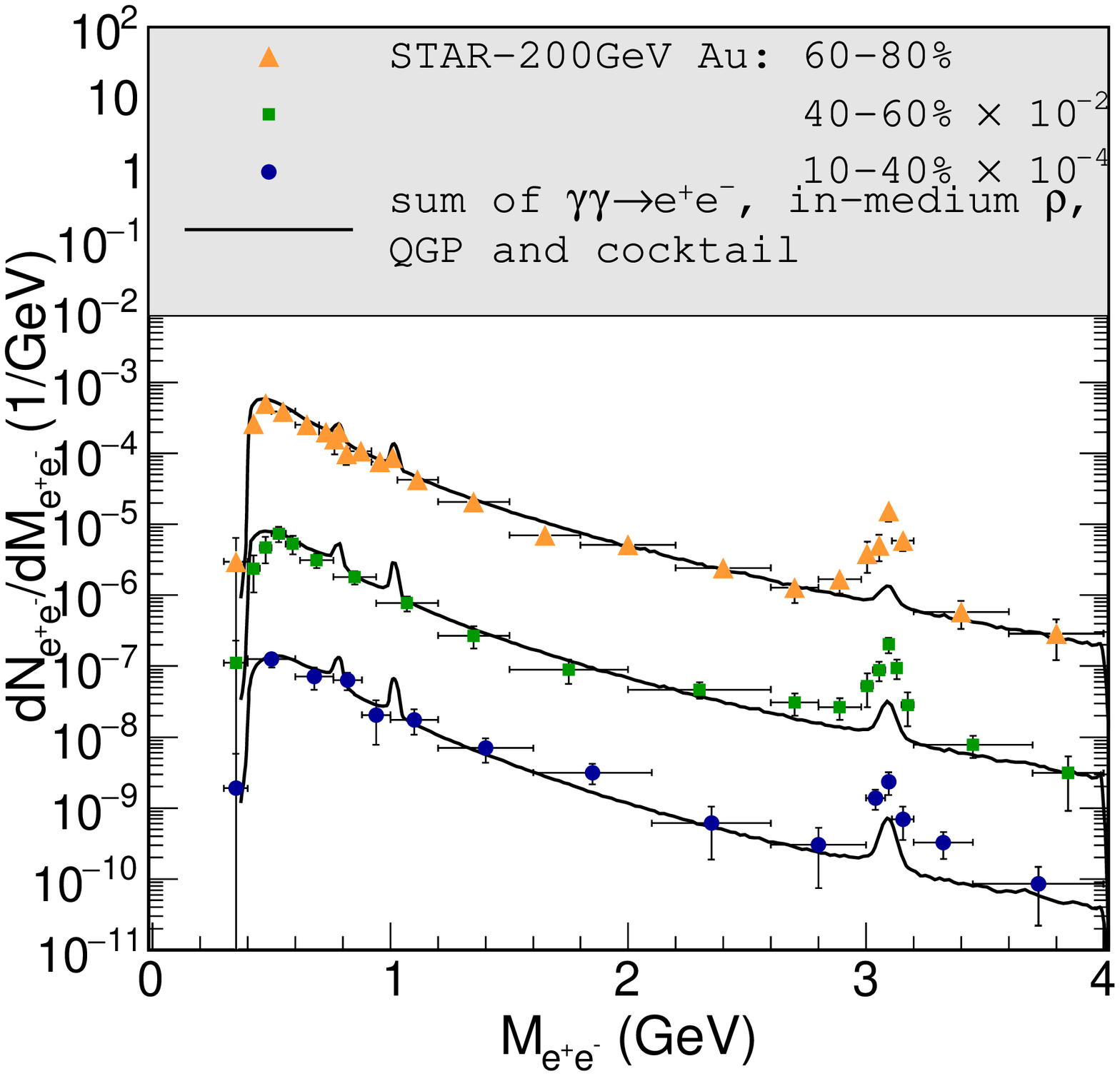}
	\caption{
		Left: Dielectron invariant-mass spectra for pair-$P_T$$<$0.15\,GeV in Au+Au($\sqrt{s_{NN}}$=200\,GeV)
		collisions for 3 centrality classes including experimental acceptance cuts ($p_t>$0.2\,GeV, 
		$|\eta_e|$$<$1 and $|y_{e^+e^-}|$$<$1) for $\gamma \gamma$ fusion (solid lines), thermal radiation 
		(dotted lines) and the hadronic cocktail (dashed lines);
		right panel: comparison of the total sum (solid lines) to STAR data~\cite{Adam:2018tdm}.
	}
	\label{fig:STAR_Mll}
\end{figure}

In Fig.~\ref{fig:STAR_Mll} we show dielectron invariant-mass 
spectra for small pair $P_{T} <$ 0.15~GeV and three different 
centrality classes as selected in the STAR analysis: peripheral (60-80\%), 
semi-peripheral (40-60\%) and semi-central (10-40\%) collisions. 
We also include the experimental acceptance cuts on the
single-lepton tracks as applied by STAR, and take the cocktail
contribution as provided by STAR~\cite{Adam:2018tdm} 
representing the final-state decays of the produced hadrons. 
In peripheral collisions the photon-photon contribution dominates while
in semi-central collisions all three contributions are of similar magnitude. 
Their sum yields a rather good agreement with the STAR data, except for the 
$J/\psi$ peak region. Our calculations only contain incoherent $J/\psi$ 
production, from binary nucleon-nucleon collisions;
we conjecture that the missing contribution is due to a coherent 
contribution~\cite{KS2016}.

\begin{figure}[!t]
\centering
	\includegraphics[scale=0.35]{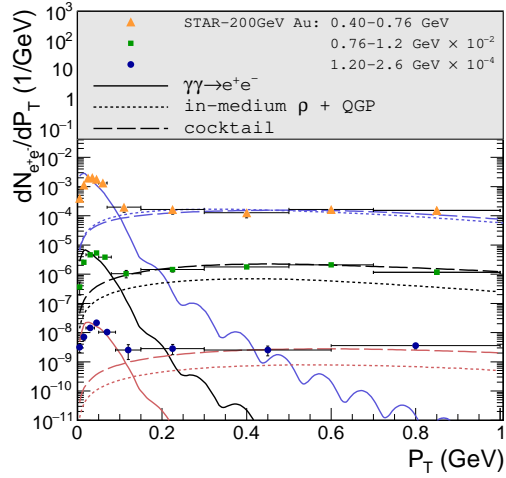}
	\caption{
		$P_T$ spectra of the individual contributions (line styles as in the previous figure)
		in 3 different mass bins for the 60-80\% centrality of the 
		Au+Au collisions ($\sqrt{s_{NN}}$=200\,GeV), compared to STAR data~\cite{Adam:2018tdm}.
	}
	\label{fig:STAR_pt}
\end{figure}

We also calculated the pair-$P_T$ distributions for the $\gamma\gamma$ 
fusion mechanism and combine it with the ones
from thermal radiation and the cocktail in Fig.~\ref{fig:STAR_pt}. 
One can clearly identify the low-$P_T$
region where the $\gamma \gamma$ fusion dominates, although the width of 
the low-$P_T$ peak in the data is slightly smaller than for the data.
In our calculations we used the realistic nuclear form factor from
Ref. \cite{Klein:2016yzr}, which leads to the oscillations in the $P_T$
distributions for the coherent photon mechanism.

\begin{figure}[!h]
	 \includegraphics[scale=0.21]{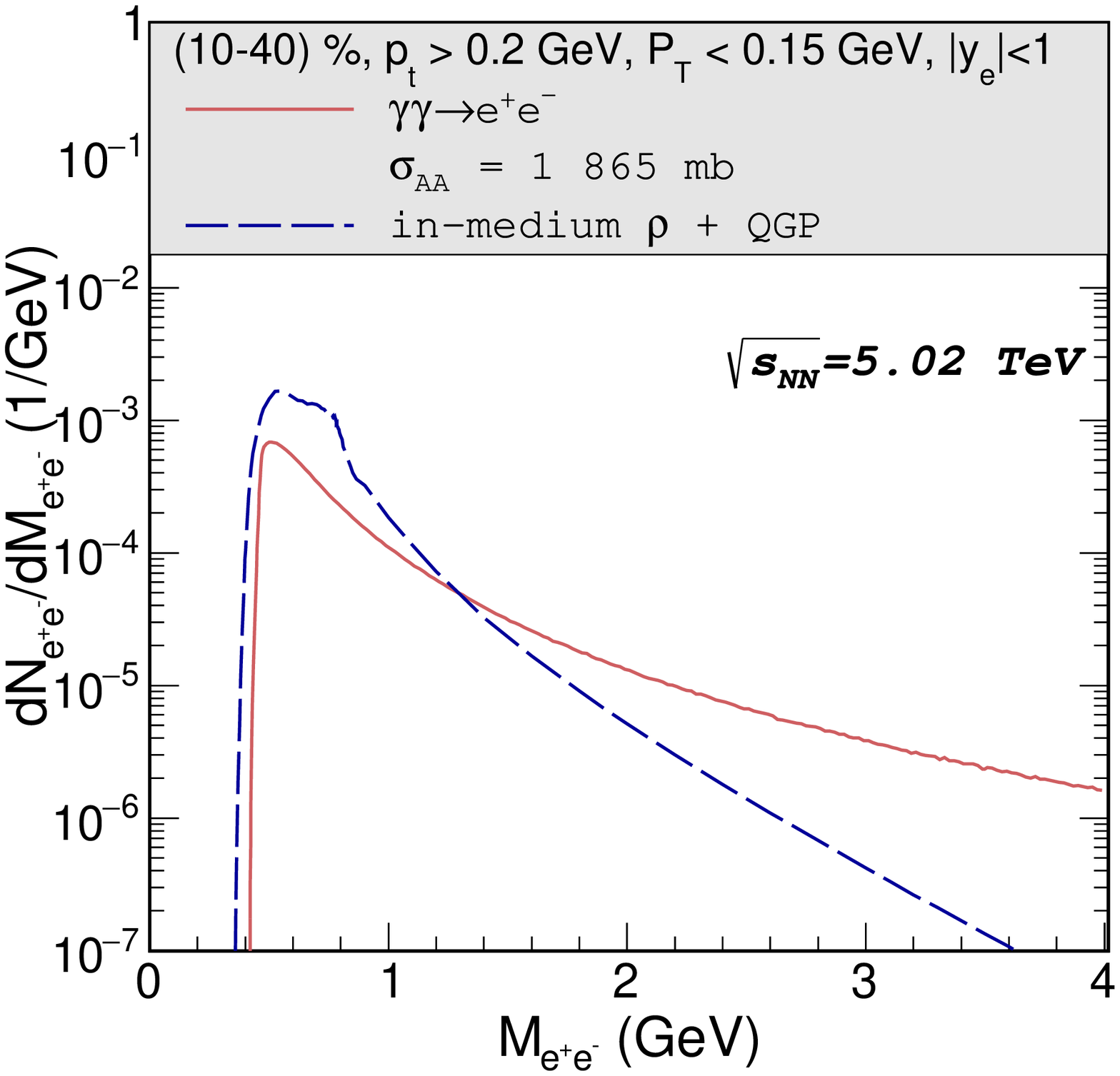}
\hspace{-0.4cm}
	 \includegraphics[scale=0.21]{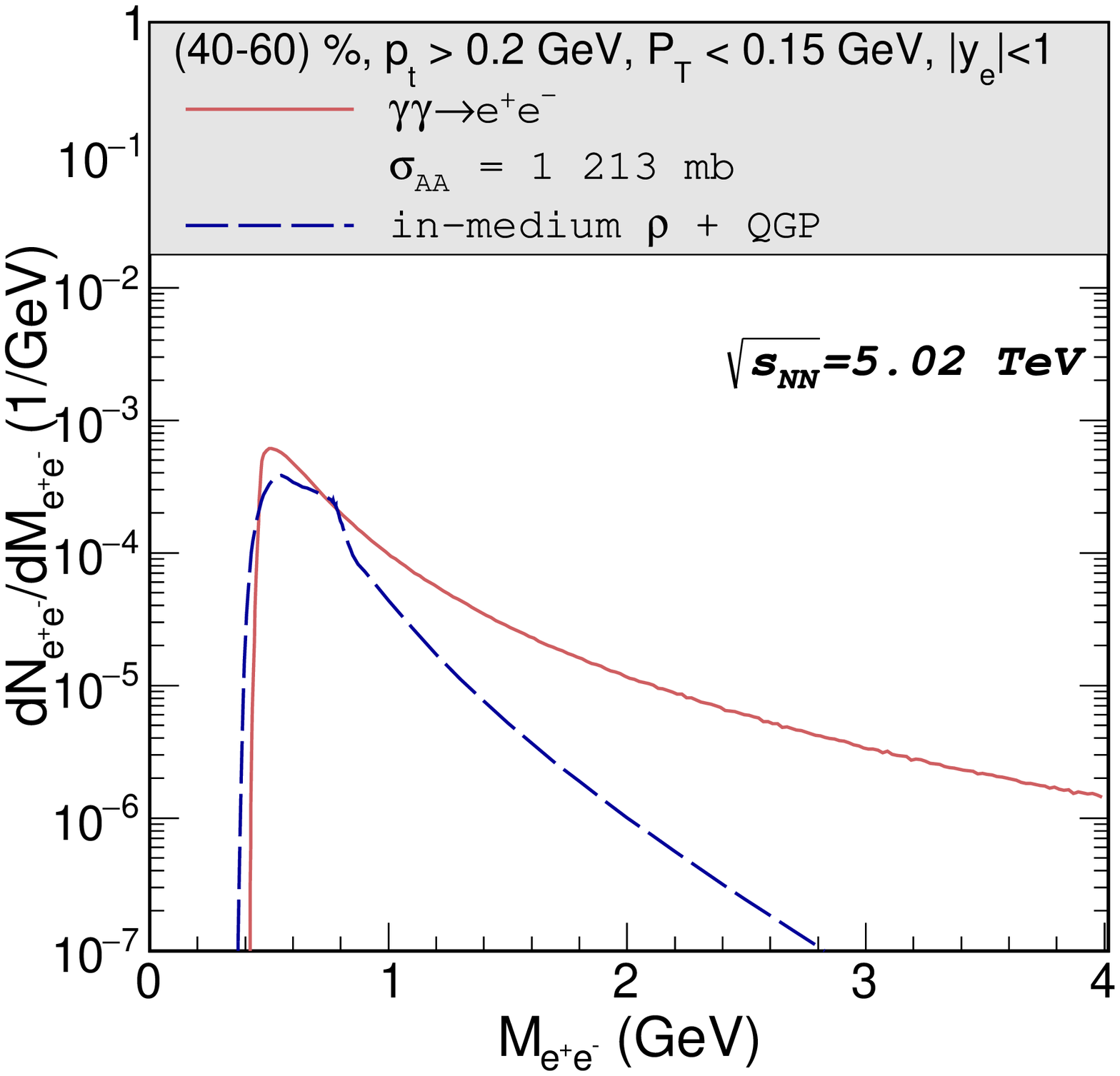}
\hspace{-0.4cm}
	 \includegraphics[scale=0.21]{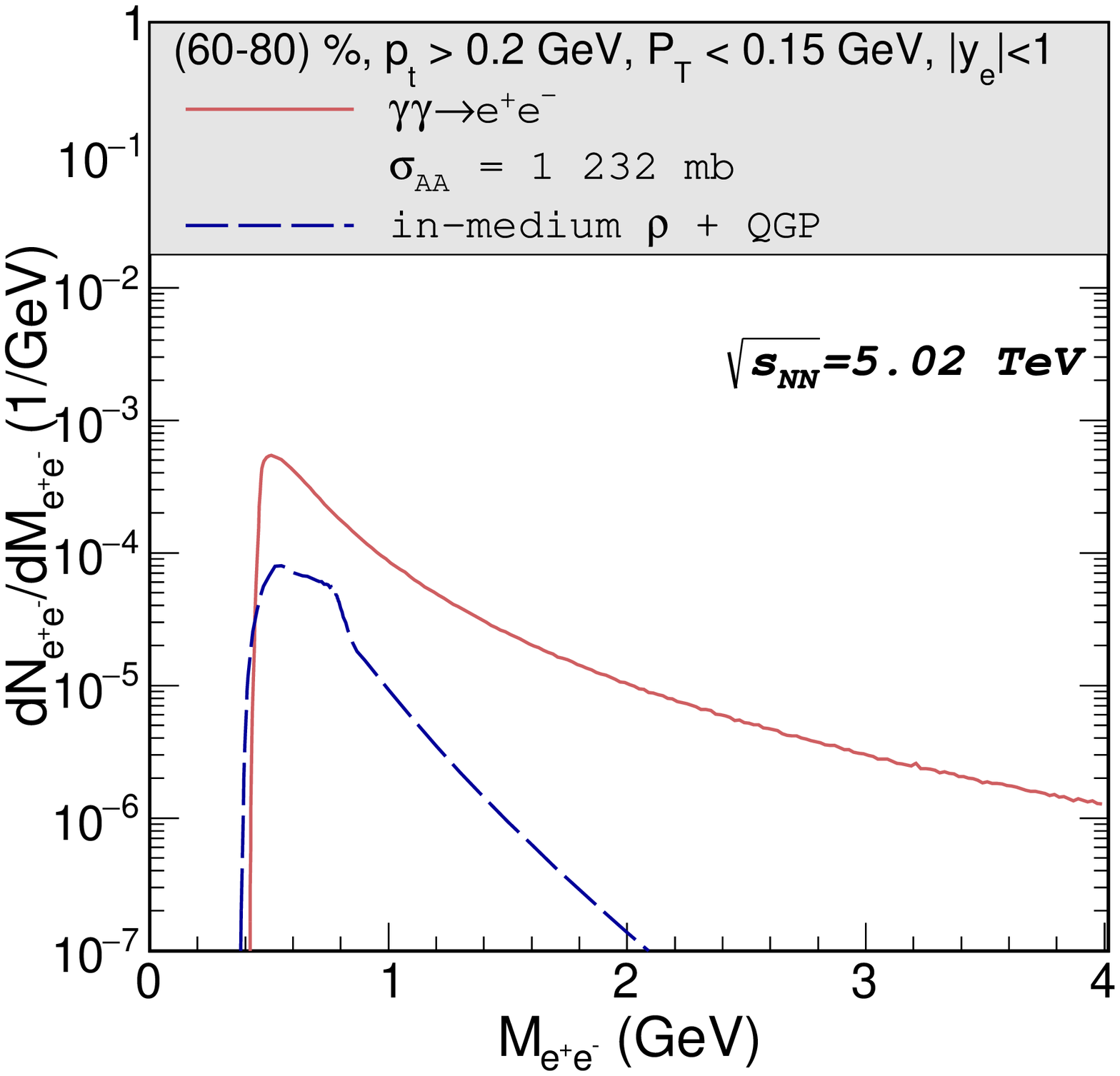}
	\caption{Our predictions for low-$P_T$ dilepton radiation in Pb+Pb ($\sqrt{s_{NN}}$=5.02\,TeV)
collisions from coherent $\gamma\gamma$ fusion (solid lines) and thermal radiation 
(dashed lines) for three centrality classes and acceptance cuts as specified in the 
figures.
	}
\label{fig:ALICE}
\end{figure}

In Fig.~\ref{fig:ALICE} where we show our predictions 
for the two sources for Pb+Pb collisions at $\sqrt{s_{NN}}$=5.02\,TeV 
for the same centrality classes and single-lepton acceptance cuts as 
for our RHIC calculations. Compared to the latter, the picture is 
qualitatively similar, although the strength of thermal
contribution is relatively stronger, especially in semi-peripheral and 
central collisions where it is comparable and even larger
than the $\gamma\gamma$ yield at low mass. 

\begin{figure}[!t]
\centering
	\includegraphics[scale=0.4]{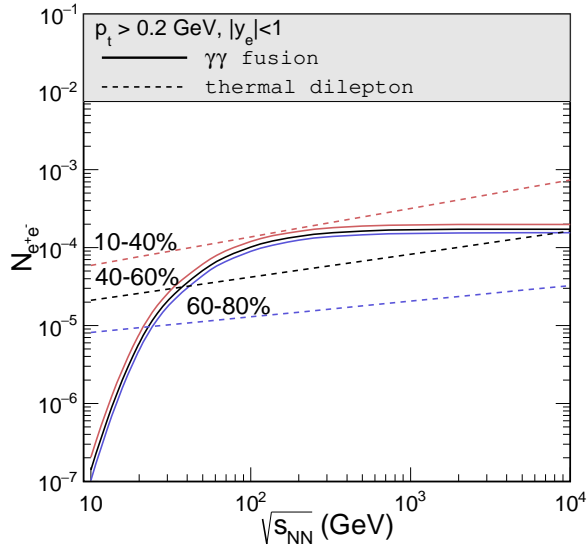}
	\caption{Excitation function of low-$P_T$ 
($<$0.15\,GeV) dilepton yields from $\gamma \gamma$ fusion (solid lines) and 
thermal radiation (dashed lines) in collisions of heavy nuclei 
(A$\simeq$200) around midrapidity in three centrality classes, 
including single-$e^\pm$ acceptance cuts.
	}
\label{fig:sig_tot_sqrts}
\end{figure}

In Fig.\ref{fig:sig_tot_sqrts} we present the dependence of the two
different contributions (total cross section) for selected centrality
classes as a function of $\sqrt{s}$. While at low energies 
the photon-fusion mechanism is negligible, it quickly rises with energy 
and saturates at about RHIC energies. In contrast, the thermal
contribution grows gradualy with energy.
This plot shows that the RHIC energy $\sqrt{s_{NN}}$ = 200 GeV is the most
favorable for observing the photon-photon fusion.

\section{Conclusions}

We have presented a theoretical study 
of photoproduction mechanism in the case when nuclei collide and 
produce quark-gluon plasma and as a consequence considerable number 
of hadrons is produced.
On theoretical side, the nuclear photoproduction in UPC was treated
in the equivalent photon approximation with photon fluxes
and photon-nucleus cross section being the basic ingredients of the
approach.

We have assumed that the whole nucleus produces photons. The photon (or
hadronic photon fluctuation) must
hit the other nucleus to produce the $J/\psi$ meson.

The question arises how to treat the region of overlapping
colliding nuclei in the impact parameter space 
where some absorption of $J/\psi$ may be expected.
We include the effect of the ''absorption'' by modifying
effective photon fluxes in the impact parameter space
by imposing additional geometrical conditions on
impact parameters (between photon and nuclei and/or 
between colliding nuclei).

As an example,  we have considered a 
vector-dominance based model which includes multiple scattering effects.
Any other model/approach can be applied in future.

By modifying standard photon fluxes valid for UPC by 
collision geometry we have calculated cross section
for different centrality bins relevant for the ALICE Collaboration
analysis. Our results have been compared with their data.
We have obtained a reasonable agreement for peripheral and semi-central
collisions and set limits for the cross section for 
the semi-central collisions.

Our lower limit is, however, somewhat model dependent.
Since in our calculations we have used coherent $\gamma A \to J/\psi A$
cross section our lower limit may be overestimated especially
for small impact parameters.

The time picture of the whole process is not clear to us in the moment.
The rather reasonable agreement of our quite simplified approach 
with the ALICE data suggest that the "coherent"
(assumed by the formula used for the $\gamma A \to J/\psi A$ process) 
scattering of the hadronic fluctuation happens before the nucleus 
undergoes the process of deterioration due to
nucleus-nucleus collision and before
the quark-gluon plasma is created. 

Here we have discussed analysis for a forward rapidity range. 
There the $J/\psi$ quarkonia are emitted forward with
large velocity therefore they could potentially escape
from being melted in the quark-gluon plasma.
At midrapidities the situation could be slightly different.
The ALICE Collaboration would repeat their analysis also in the
midrapidity range and verify the $p_t \approx$ 0 enhancement.

We have discussed also low-$P_T$ dilepton production in ultrarelativistic
heavy-ion collisions, 
by conducting systematic comparisons of the two sources of dileptons
The former was taken from a model including in-medium 
hadronic and QGP emission rates, while the latter was calculated
utilizing photon fluxes with realistic nuclear form factors including 
the case of nuclear overlap. We have found 
that the combination of the two sources 
(augmented by a contribution from the hadronic final-state decay cocktail) 
gives a good description of low-$P_T$ dilepton data in
Au-Au ($\sqrt{s_{NN}}$=200\,GeV) collisions in three centrality 
classes for invariant masses from threshold to 4\,GeV 
(with the exception of the $J/\psi$ peak related to coherent production
of $J/\psi$). 
The coherent emission of $e^+ e^-$ pairs was found to be dominant for 
the two peripheral samples, and comparable to the cocktail 
and thermal radiation yields in semi-central collisions. 

At high-energies, the situation is 
similar to RHIC energies.
The interplay of these processes at the LHC is of particular interest 
in view of plans by 
the ALICE collaboration~\cite{Antinori:2018} to lower the
single-electron $p_t$ cuts and measure very-low mass spectra.

We have summarized our results in an excitation function of low-$P_T$ 
radiation covering  three orders of magnitude in collision energy. 
While coherent production increases rather 
sharply, and then levels off, near $\sqrt{s_{NN}}$$\simeq$100\,GeV, 
thermal radiation increases more gradually with $\sqrt{s_{NN}}$. 
This explains why the latter is dominant at the SPS, the former 
dominates at RHIC, and the latter becomes more important again at the LHC.

\vspace{0.5cm}

{\bf Acknowledgement}

I am indebted to Mariola K{\l}usek-Gawenda, Ralf Rapp and Wolfgang
Sch\"afer for collaboration on the issues presented here.



\begin{thebibliography}{100}

\bibitem{ALICE2016}
E.L. Kryshen for the collaboration [ALICE Collaboration],
Nucl. Phys. \textbf{A967} (2017) 273. 

\bibitem{KS2016}
M. K{\l}usek-Gawenda and A. Szczurek,\\
``Photoproduction of $J/\psi$  mesons in peripheral and semicentral 
heavy ion collisions'',\\
Phys. Rev. {\bf C93} (2016) 044912.

\bibitem{Adam:2018tdm}
J.~Adam {\it et al.} [STAR Collaboration],
Phys. Rev. Lett. \textbf{121} (2018) 132301.

\bibitem{KRSS2019}
M. K{\l}usek-Gawenda, R. Rapp, W. Sch\"afer and A. Szczurek,\\
"Dilepton Radiation in Heavy-Ion Collisions at Small Transverse
Momentum",\\
Phys. Lett. {\bf B790} (2019) 339. \\

\bibitem{Bertulani:1987tz}
C.~A.~Bertulani and G.~Baur,
Phys.\ Rept.\  {\bf 163} (1988) 299.

\bibitem{Baur:2001jj}
G.~Baur, K.~Hencken, D.~Trautmann, S.~Sadovsky and Y.~Kharlov,
Phys.\ Rept.\  {\bf 364} (2002) 359.

\bibitem{Klein:2016yzr}
S. R. Klein, J. Nystrand, J. Seger, Y. Gorbunov and J. Butterworth,
Comput. Phys. Commun. \textbf{212} (2017) 258.

\bibitem{Antinori:2018}
F.~Antinori, P.~Braun-Munzinger and S. Fl\"orchinger, priv. comm. (2018).

\end{thebibliography}
\end{document}